# Is scientific literature subject to a 'sell-by-date'?
## A general methodology to analyze the 'durability' of scientific documents

Rodrigo Costas[1], Thed N. van Leeuwen, and Anthony F.J. van Raan

*Centre for Science and Technology Studies*
Leiden University
Wassenaarseweg 62A
P.O. Box 905
2300 AX Leiden
The Netherlands

*Abstract*

*The study of the citation histories and ageing of documents are topics that have been addressed from several perspectives, especially in the analysis of documents with "delayed recognition" or "sleeping beauties". However, there is no general methodology that can be extensively applied for different time periods and/or research fields. In this paper a new methodology for the general analysis of the ageing and "durability" of scientific papers is presented. This methodology classifies documents into three general types: Delayed documents, which receive the main part of their citations later than normal documents; Flash in the pans, which receive citations immediately after their publication but they are not cited in the long term; and Normal documents, documents with a typical distribution of citations over time. These three types of durability have been analyzed considering the whole population of documents in the Web of Science with at least 5 external citations (i.e. not considering self-citations). Several patterns related to the three types of durability have been found and the potential for further research of the developed methodology is discussed.*

**Introduction**

One topic of high concern of researchers is how many years they need to wait until their papers can be properly acknowledged and accepted (and therefore cited) by their scientific community. Previous authors have studied the effects of the "scientific prematurity" (Stent, 1972) or "delayed recognition" (Garfield, 1980) related to all cases of papers that have been cited later in comparison with the average papers in their research field.

The effects of the ageing of scientific publications are important for the study of scientific communication (Pollmann, 2000). Policy makers and research managers are often inclined to think that science is primarily a 'hectic' business in which research results will prove their impact soon after publication. This attitude underestimates the crucial importance of durability of research results. Therefore the study of the ageing and delayed reception of scientific documents is certainly also an important topic for the application of bibliometrics in research assessment, for instance, the correct establishment of citation windows and proper calculation of impact indicators.

---
[1] Corresponding author, e-mail address: costascomesanar@cwts.leidenuniv.nl

The study of ageing and citation histories of documents has been addressed from several perspectives (Aversa, 1985; Glänzel & Schoepflin, 1995; Moed et al, 1998; Aksnes, 2003), especially the analysis and detection of documents with "delayed recognition" (Garfield, 1980; Glänzel et al, 2003) or even "Sleeping Beauties" (van Raan, 2004). The opposite effect of delayed recognition was also described as "Flash in the pans" (van Dalen & Henkens, 2005): documents that are noticed immediately after publication and frequently cited but which do not seem to have a lasting impact and die early in life. Both aspects (delayed recognition and flash in the pans) deal with the more general ideas on durability and obsolescence of documents, as well as with the ageing of scientific literature, which is strongly related to the impact of documents over time.

Earlier studies have focused mainly on the ageing and citation patterns of highly cited papers (Cano & Lind, 1991; Aversa, 1985; Aksnes, 2003; Levitt & Thelwall, 2008, 2009) trying to establish their main determinants and properties. However, until now there is no proper methodology for a global durability analysis of documents regardless degree of citedness, year of publication or research field. Previous attempts were based on fixed years of publication and the thresholds for the classification of documents were based on the selection of the authors (Aksnes, 2003; Moed et al, 1998). Therefore, the proposal of an integral methodology with classification of documents according to their durability regardless of publication year and/or total number of citations is necessary, as well as a general "technical" definition of different types of durability of documents.

In this paper we propose a methodology to classify documents according to the "Durability" of their citations. Here "Durability" is understood as the different patterns of the citation history of documents. The importance of such a methodology for the identification of papers deviating from the typical citation patterns was already suggested by Garfield (1980), but until now there is no general methodology that can be used for separate years, disciplines and document types.

**Objectives**

Our main objective is to develop a general methodology for the classification of research publications according to the "durability" of their citations. We aim at a classification of all documents regardless year of publication or degree of citations (i.e., taking into account all and not only highly cited documents). This methodology considers the different Web of Science[2] Journal Subject Categories[3] (fields) where documents are classified.

After the development of the methodology, we analyze the main characteristics of documents according to their durability. The analysis focuses on a general picture of the different types of durability and how citations histories of documents evolve over time. This provides us with a general description and particularly citation patterns for the different types of durability that can be used in the analysis of different fields as well as of different citation windows.

---

[2] Thomson Reuters is the producer and publisher of the Web of Science (WoS) that covers the (former) Science Citation Index (-extended), the Social Science Citation Index, and the Arts & Humanities Citation Index. Throughout this paper we use the term 'WoS' for the above set of databases.

[3] We use the definition of fields based on a classification of scientific journals into *categories* developed by Thomson Reuters. Although this classification is not perfect, it provides a clear and 'fixed' consistent field definition suitable for automated procedures within our data-system.

**Methodology**

*Development of a general methodology for the study of durability*

Our methodology for the classification of the durability of scientific papers aims at the classification of documents according to their citation histories in three general types:

- *Normal-type*: documents with the typical distribution in their citations over time, i.e., reaching a maximum in, say, three to four years after publication and then followed by an exponential decay;

- *Flash in the pans-type*: documents that tend to receive citations immediately after their publication but are not cited in the longer term;

- *Delayed-type*: documents that receive the main part of their citations later than normal documents; "Sleeping beauties" are included within this type.

This classification is similar to that used by van Dalen & Henkens (2005), however they called the third type "Sleeping beauties". We prefer the term "Delayed" because Sleeping Beauties are a rare and rather extreme phenomenon in science (van Raan, 2004). Other authors also observed similar classifications. For example, Aversa (1985) found two general patterns: "Delayed rise - Slow decline" that coincide with our Delayed-type, and "Early rise - Rapid decline" that are basically our Flash in the pans. Aksnes (2003) also considered a third type "Medium rise - Slow decline" that can be considered as strongly related to our Normal-type documents.

This paper is based on the data of documents covered by the *WoS* including more than 8,700 journals and covering all scientific disciplines (Costas & Iribarren-Maestro, 2007), thus being one of the most important databases for bibliometric studies and research assessment purposes. We have also focused on "external citations", i.e., citations received by documents after the exclusion of self-citations[4] (Costas et al, 2008). We take the position that this type of citations represents the real impact and transfer of knowledge of documents beyond their original producers.

Our approach, based on the distribution of the percentage of citations that documents receive each year (citation history), is composed of the following steps:

1. For each document in the *WoS*, its *citation history* over time is analyzed (self-citations excluded). This provides us with the evolution of citations since the year of publication until the last year considered in the analysis, see example in Table 1.

---

[4] A citation is a self-citation if any of the authors of the citing paper is also an author of the cited paper.

**Table 1**: *Example citation history document A*

| Document | Year Pub. | Tot. Ext. Cit. | 2002 | 2003 | 2004 | 2005 | 2006 | 2007 | 2008 |
|---|---|---|---|---|---|---|---|---|---|
| A | 2002 | 25 | 0 | 3 | 6 | 7 | 4 | 3 | 2 |

*According to Table 1, document A was published in 2002, received a total of 25 external citations until 2008, and the evolution of the citations is shown in the following years.*

2. For all documents with at least one citation, the percentage of citations received each year has been calculated on the basis of their citation history, as well as the cumulative value of the percentage of citations, see example in Table 2. A similar approach, calculating the yearly percentage of citations, was also used by Aksnes (2003).

**Table 2**: *Example percentage of citations and cumulative values of document A*

| Document | Year Pub. | Tot. Ext.Cit. | 2002 | 2003 | 2004 | 2005 | 2006 | 2007 | 2008 |
|---|---|---|---|---|---|---|---|---|---|
| A | 2002 | 25 | 0 | 3 | 6 | 7 | 4 | 3 | 2 |
| % cit. per year | | 100 | 0.00 | 12.00 | 24.00 | 28.00 | 16.00 | 12.00 | 8.00 |
| Cum. % | | 100 | 0.00 | 12.00 | 36.00 | 64.00 | 80.00 | 92.00 | 100.00 |

*In Table 2, we see the evolution of the percentage of citations for document A as well as its cumulative value.*

3. For each document we identified the year after publication in which the document received for the first time *at least 50%* of its citations ("Year 50%") during the time period up to and including 2008. This provides us with an ordinal number for the document (e.g., for documents published in 1994 the number 1 is given when documents received at least 50% of their citations in 1994; the number 2 when 50% of citations is reached in 1995; 3 if this happens in 1996, etc.). In our example for document A, this year is 2005 (shaded in Table 2), what means that in our example the value for "Year 50%" of document A is 4.

4. For the whole population of documents (with at least one external citation), taking into account all document types and considering the different research fields (i.e., WoS Journal Subject Categories, JSC), we calculated for all documents *of the same year of publication*, the percentiles 25 and 75 of the distribution function of the value of the new indicator "Year 50%". As noticed by Glänzel & Schoepflin (1994), the ageing behavior of documents is influenced by field characteristics, and therefore we consider the scientific field as the best reference for the classification of each document.

Taking again our example, document A is classified in one field (JSC) with P25 and P75 scores as given in Table 3.

**Table 3**: *Example of document A and the P25 and P75 of its field*

| Document | Year Pub. | Year 50% | P25-JSC | P75-JSC |
|---|---|---|---|---|
| A | 2002 | 4 | 3 | 6 |

*According to Table 3, document A is classified in a field where the P25 and 75 of the "Year 50%" of all documents published in 2002 is 3 and 6 respectively.*

5. The general criterion for the classification of documents is now as follows:

   a. Flash in the pans: the value for "Year 50%" is smaller than the P25 value for that field (<P25);
   b. Delayed documents: the value for "Year 50%" is larger than the P75 value for that field (>P75);
   c. Normal documents: the value for "Year 50%" is between the P25 and the P75 value for that field (≥P25 and ≤P75).

This criterion is based on a personal communication of Derek de Solla Price to Aversa (1985) in which he claimed that "papers exhibit three basic citation patterns" distributed among "25 per cent of papers cited at a constant rate without declining" (roughly the equivalent of our Delayed type); "25 per cent of gradually increased in citedness and then declined at a similar rate" (similar to our Flash in the pans), and finally "50 per cent cited at a constant rate for several years" (the equivalent of our Normal-type). In Fig. 1 a general scheme of the criteria proposed is shown.

**Figure 1**: *General scheme of classification of papers by Durability type*

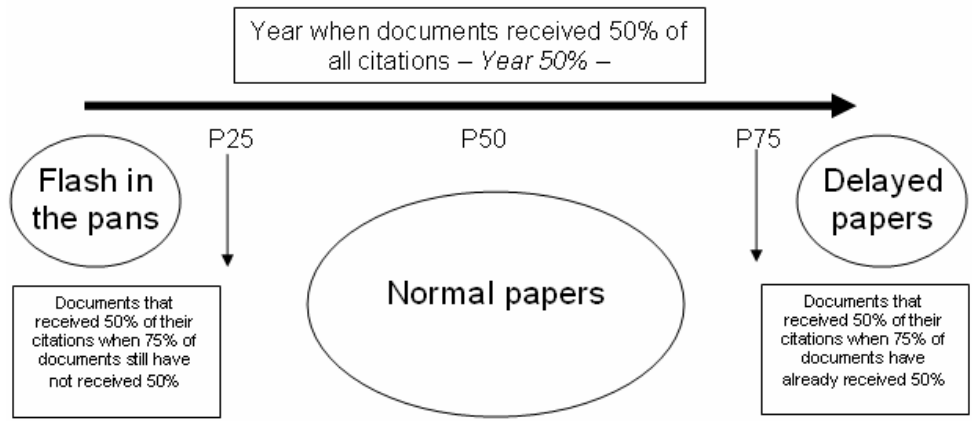

According to the above criteria Flash in the pans can then be defined as those documents that have received 50% of their citations when the 75% of other documents still have not received 50% of their citations. Normal documents are all documents that receive the 50% of their citation around year of P50 (between P25 and P75). Finally, Delayed documents are those papers that have received 50% of their citations after P75 years in their fields. Notice that the criteria of step 5 tend to favor Normal documents, because we assumed that this pattern should be preferred instead of the other two when they are equal to the thresholds (P25 and P75). Thus, in a hypothetical situation of all documents gathering 50% of their citations in the same number of years after publication, P25 and P75 are equal to P50, and according to our classification all documents would be classified as Normal.

This methodology is originally based on the JSC distribution of documents according to the classification of journals in the Web of Science (more specifically, Thomson Reuters' Journal Citation Reports). However, journals can be classified into more than one field (JSC). Consequently, documents in such journals will be assigned to more than one field and therefore to more than just one Durability type: a document can be in one of the categories a "Normal" document, and in another category "Delayed". To avoid this "document schizophrenia" and to characterize each document with just one single type, the following steps are proposed:

6. For documents in more than one field (i.e., more than one JSC) and with more than one "Durability class", different "Class-values" are assigned to different types:

    a. Flash in the pans  → 1
    b. Normal            → 2
    c. Delayed           → 3

7. The number of different fields (JSCs) of each document is also calculated and the total durability class-values are summed. In Table 4 the example of document A is presented. This document is classified in two fields (Entomology and Biochemistry) with different P25 and P75 and different Durability types.

**Table 4**: *Document A in more than one field*

| Document | Year pub. | Year 50% | JSC | P25 | P75 | Durability type | Class-value |
|---|---|---|---|---|---|---|---|
| A | 2002 | 4 | Entomology | 3 | 6 | Normal | 2 |
|   |      |   | Biochemistry | 2 | 3 | Delayed | 3 |

*We see that document A, assigned to two different fields, has two different durability types in these fields and thus one different class-value for each field.*

8. Finally, the total durability class-values of each document are divided by the number of fields. Thus, all documents get values between 1 and 3. The final criteria for classification is as follows:

    a. Flash in the pans → ≤1.5
    b. Normal            → > 1.5 and < 2.5
    c. Delayed           → ≤2.5

This second classification is based on the idea that the most frequent Durability type among the different fields of papers is the most important. This means that in this second classification implicitly deviant types (i.e., Flash in the pans and Delayed) prevail when they appear with the same frequency as the Normal type, but they are mutually balanced when they appear together in the same document, in favor of the Normal type. In other words, a paper assigned to two fields and being a Delayed in the first field and a Flash in the pan in the other, is finally considered as Normal). Further analyses and refinements of this aspect of the methodology will be developed in the next stage of this study.

**Table 5**: *Final Durability type for document A*

| Document | Year pub. | Year 50% | Tot. JSC | Tot. Sum. Class-values | Avg. Class-value | Final Durability type |
|---|---|---|---|---|---|---|
| A | 2002 | 4 | 2 | 5 | 2.5 | Delayed |

As a final result of this methodology, all documents are classified in just one final Durability type: in our example in Table 5, document A gets a final classification as Delayed, which makes the analysis of all documents more straightforward. Thus, with our methodology we are able to classify all documents regardless of their degree of citedness (even documents with just one external citation can be classified depending on the year they receive their citation), taking into account both their year of publication and their (if this is the case) different fields.

**Results**

The above discussed methodology has been applied to all documents in the *Web of Science* database published between 1980 and 2008 (in total 30,445,406 documents), including all document types, languages, years of publication, etc. Essential elements of the analysis to identify the durability type of each publication are year of publication, citation trend up to and including 2008, and field (JSC). In this section we discuss our results concerning the main characteristics of the three types of Durability.

First we present a general description of the three types of durability and their properties. This analysis provides an impression of the documents behind the three types. Although all documents with at least one external citation were identified in the WoS, we considered in our analysis only those publications with a minimum of 5 external citations and published between 1980 and 2003 (8,340,513 documents) in order to avoid the influence of hardly cited documents and also documents published at the very end of the period. Thus, we have 'source publications' -which are the cited publications- for the period 1980-2003, and for the citing publications -in order to count citations- we use the period 1980-2008.

*1. General properties of documents according to their durability*

In Table 6 we present a general description of documents, considering different bibliometric properties of documents classified in the three Durability classes.

**Table 6**: *General values for documents in different durability types*

| Durability | Total pub. | % | Authors/doc | Instit/doc | Countries/doc | Kwords/doc | Pages/doc | Refs/doc |
|---|---|---|---|---|---|---|---|---|
| Delayed | 1,665,712 | 19.97 | 3.37±3.41 | 1.99±1.36 | 1.14±0.44 | 4.82±1.80 | 9.22±37.05 | 28.56±28.82 |
| Flash in the pans | 814,098 | 9.76 | 3.74±9.57 | 2.07±1.77 | 1.16±0.54 | 4.86±1.79 | 8.11±39.69 | 29.31±30.09 |
| Normal | 5,860,703 | 70.27 | 3.75±6.48 | 2.25±1.66 | 1.18±0.53 | 4.91±1.82 | 9.29±42.66 | 32.96±33.18 |

Note: values are indicated by the *mean* value ± standard *deviation*.
"Kwords/doc" refers to the number of author-given keywords per document.

Statistically significant differences (p<0.000) have been observed among all Durability types in all indicators (test U-Mann Whitney). We also observed significant differences for the direct comparison of Delayed and Flash in the pans (p<0.000). A further important interesting observation is that the distribution of documents among the three types does not come up to a 25-50-25 distribution. The reason for this is that we discarded very lowly cited documents and that the "threshold equals" as discussed in Step 5 tends to favor "Normal" documents. According to Table 6, Delayed documents have a slightly lower number of authors per document, lower number of institutes and countries per document, and they also have fewer author-keywords and references per document.

On the other hand, comparing Flash in the pans with Delayed documents, we observe that the first are generally documents with more authors, more countries and more institutes per document than the Delayed documents. They also have more references per document than Delayed. It is important to stress that Flash in the pans are on average the shortest documents (i.e., less pages per document). The distribution of all three Durability types by document type was also analyzed from a general perspective; see Table 7 and Figure 2.

**Table 7**: *Distribution of document types among durability types*

| Durability | Doc. Type | Tot. Docs. | % |
|---|---|---|---|
| Delayed | Articles | 1,496,444 | 89.84 |
| | Notes | 74,104 | 4.45 |
| | Reviews | 52,341 | 3.14 |
| | Letters | 23,707 | 1.42 |
| | Editorials | 10,950 | 0.66 |
| | Others | 8,166 | 0.49 |
| Flash in the pan | Articles | 677,461 | 83.22 |
| | Notes | 39,035 | 4.79 |
| | Reviews | 29,713 | 3.65 |
| | Letters | 20,454 | 2.51 |
| | Editorials | 28,708 | 3.53 |
| | Others | 18,727 | 2.30 |
| Normal | Articles | 5,202,724 | 88.77 |
| | Notes | 205,920 | 3.51 |
| | Reviews | 256,280 | 4.37 |
| | Letters | 84,354 | 1.44 |
| | Editorials | 74,063 | 1.26 |
| | Others | 37,362 | 0.64 |

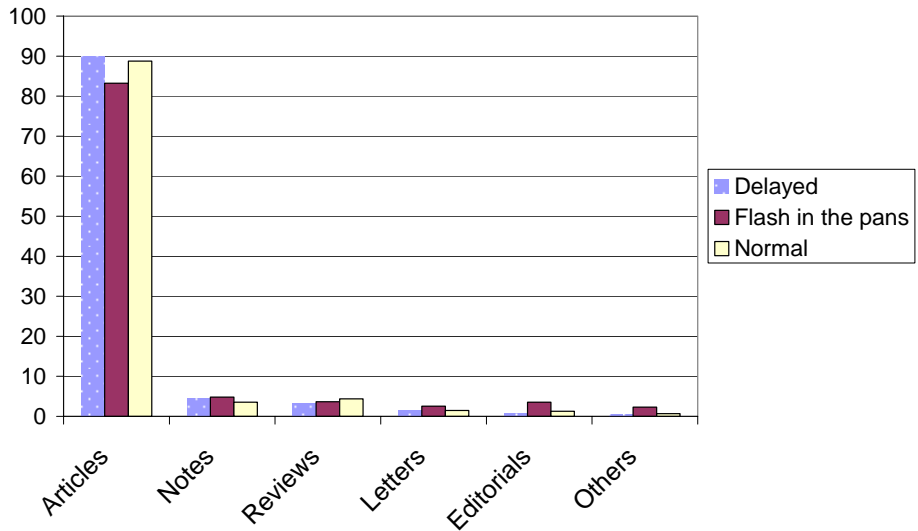

**Figure 2**: *Durability classes by document types*

We immediately observe that Flash in the pans are published proportionally more as Notes, Letters, Editorials, etc. than the other two types. On the other hand, Delayed documents are more represented in the document type "Articles". These results already provide a first explanation why Flash in the pans are generally shorter documents, because this type contains proportionally more Notes, Letters and Editorials than the other two. In this line, a closer look at two 'letter' journals (which include mainly rapid communications; Moed et al, 1998) such as *Science* and *Nature*, shows that these journals cover 10.9% and 10.5% of Flash in the pans respectively. This means that they publish proportionally more papers of this type than the observed average for the whole database (9.8%). These results corroborate previous results by Glänzel & Schoepflin (1994, 1995) that document and journal types influence the ageing behavior of documents. This implies that letters, short communications and occasionally short review articles represent a much faster communication, in other words, a higher degree of Flash in the pans.

The performance of the three types of durability was also studied on the basis of the CWTS standard indicators[5] (Moed et al, 1995). For the analysis presented in this paper, only Articles, Notes, Letters and Reviews were considered (Table 8). A variable citation window was used, which means that citations are counted from the year of publication of documents up to and including 2008.

---

[5] We stress again that self-citations have been removed from the publications of the individual researchers but also from all publications used as an international reference. Thus, only 'external citations' (i.e., citations given by authors different from the co-authors of the original paper) have been considered for the calculation of all indicators.

**Table 8**: *Standard indicators by Durability type*

| Durability | Period citations | P | C | CPP(*) | FCSm | JCSm | %sc | CPP/FCSm | CPP/JCSm | JCSm/FCSm |
|---|---|---|---|---|---|---|---|---|---|---|
| Delayed | 1980-08 | 1,646,596 | 56,258,641 | 34.17 | 17.71 | 21.42 | 0.12 | 1.93 | 1.60 | 1.21 |
| Flash in the pan | 1980-08 | 766,663 | 10,892,718 | 14.21 | 18.49 | 22.31 | 0.19 | 0.77 | 0.64 | 1.21 |
| Normal | 1980-08 | 5,749,278 | 169,749,327 | 29.53 | 17.81 | 24.42 | 0.14 | 1.66 | 1.21 | 1.37 |

Note: period 1980-2003 for documents and 1980-2008 for citations. For an explanation of the indicator symbols, see the text box below this table.
(*) The total number of external citations of documents in each durability type have also been analyzed through the Mann-Whitney test which indicated significant differences among the three types of durability (p<0.000).

---

**Standard Bibliometric Indicators:**

- Number of publications *P* in WoS-covered journals;
- Number of citations *C* received by *P* during the specified period, *without* self-citations;
- Average number of citations per publication, without self-citations (***CPP***);
- Journal-based worldwide average impact as an international reference level (***JCS***, journal citation score), without self-citations; as many journals are involved in the different classes, we use the average ***JCSm***; for the calculation of ***JCSm*** the same publication and citation counting procedure, time windows, and article types are used as in the case of ***CPP***;
- Field-based worldwide average impact as an international reference level (***FCS***, field citation score), without self-citations; as many fields are involved, we use the average ***FCSm***; for the calculation of ***FCSm*** the same publication and citation counting procedure, time windows, and article types are used as in the case of ***CPP***; we refer in this article to the ***FCSm*** indicator as the 'field citation density';
- The percentage of self-citations ***%sc*** in total number of citations;
- Comparison of the ***CPP*** with the world-wide average based on ***JCSm*** as a standard, without self-citations, indicator ***CPP/JCSm***;
- Comparison of the ***CPP*** with the world-wide average based on ***FCSm*** as a standard, without self-citations, indicator ***CPP/FCSm***, this is our 'crown indicator';
- Ratio ***JCSm/FCSm*** is the relative, field-normalized journal impact indicator.

---

According to Table 8, Delayed documents present the highest average impact in citations per paper (***CCP***) compared to the other types of documents. However they are published in fields (***FCSm***) and journals (***JCSm***) with the *lowest* level of citation density as compared to the other durability types. Also for the (long life) field-specific impact ***CPP/FCSm*** they show the highest level, as well as the highest ***CPP/JCSm***. Another interesting finding is that Delayed documents relate to the lowest percentage of self-citations. This finding is in agreement with the general observation that the share of self-citations is decreasing as a function of time, in other words, the longer after publication we measure the total number of citations, the smaller the fraction of self-citations.

Flash in the pans show the lowest ***CPP***, but these documents are normally published in fields with high citation density (***FCSm***), although their journals are not cited at the same level (intermediate ***JCSm***). They present the lowest ***CPP/FCSm*** and ***CPP/JCSm*** and the highest percentage of self-citations.

### *2. Effects of the durability on citation windows*

We also analyzed the effect of different citation windows for calculation of indicators. First, the evolution of the ***CPP/FCSm*** of the three different durability types has been analyzed with variable citation windows. For better understanding of the results, only documents published in 1981 were considered in this analysis, see Fig. 3.

**Figure 3**: Evolution of *CPP/FCSm* with increasing citation window

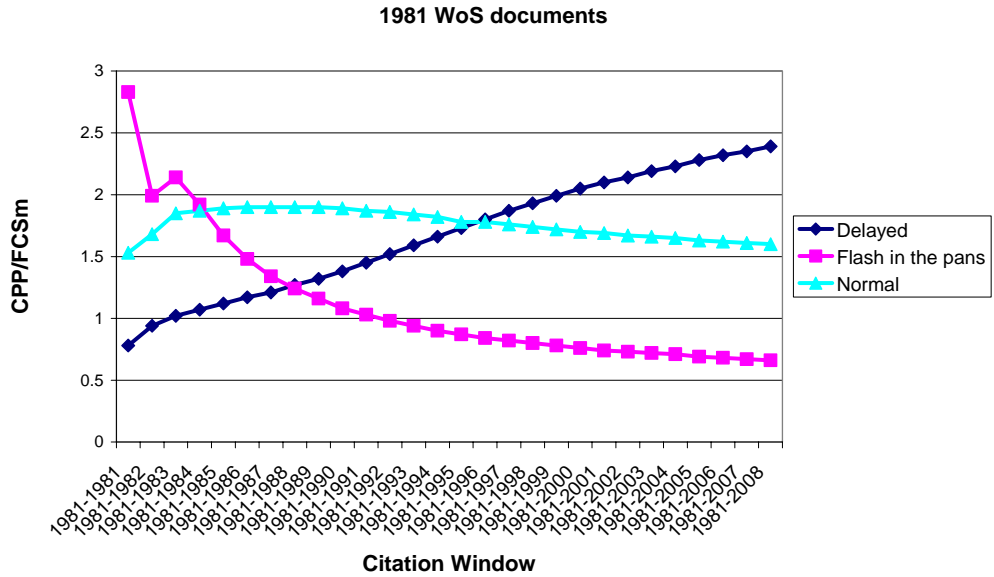

Fig. 3 presents the evolution of **CPP/FCSm** for documents published in 1981 with a yearly increasing citation window. The graphs clearly show that in case of a lengthening of the citation window, the **CPP/FCSm** values of the different durability types follow quite different patterns. More specifically, Flash in the pans have the highest scores with the shortest citation window, but they display a *decreasing* pattern as the citation window becomes longer. Delayed documents have the opposite pattern, an increasing citation window implies that their **CPP/FCSm** also increases over time, and this trend holds during the entire period. Finally, Normal documents show a quite stable pattern, increasing during 3 to 4 years after publication and slightly decreasing (but quite stable) afterwards.

According to these findings it is clear that the selected citation window has an effect on the final results. It is also evident that this first indication of differences between the three types of durability as a consequence of changing citation patterns is very relevant for the use of bibliometric indicators in research assessment procedures. To test this effect, different *fixed* citation windows (3, 5, 10, and 20 years) have been considered for the documents in the three durability types. In Fig. 4 the evolution of the **CPP/FCSm** with different fixed citation windows is presented for the three types of durability.

**Figure 4**: Evolution of *CPP/FCSm* with different fixed citation windows

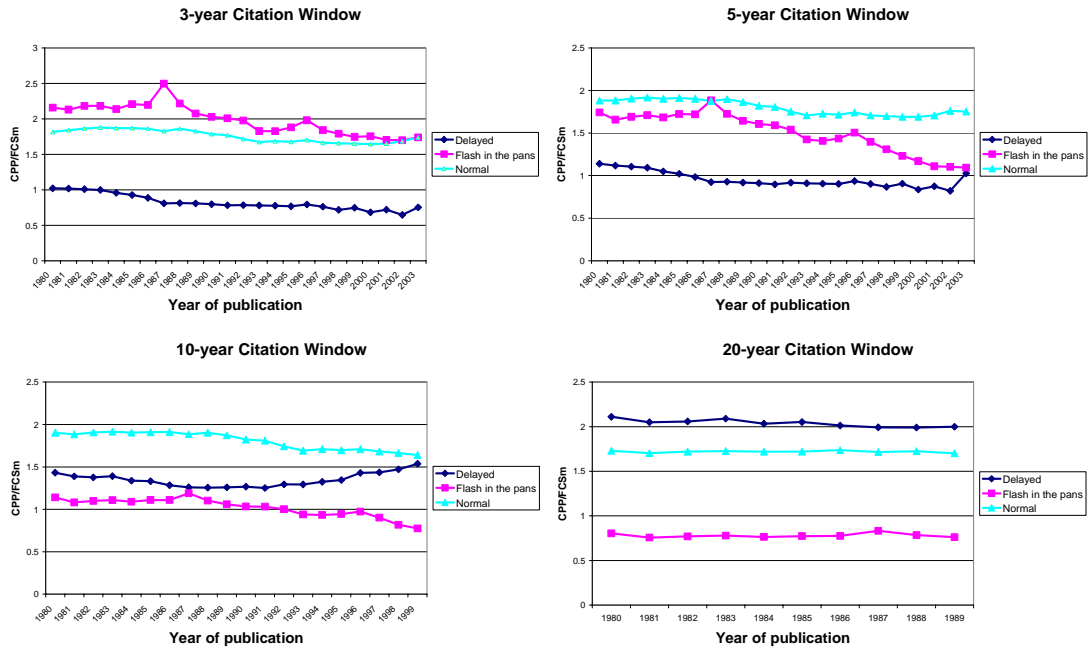

With the shortest citation window -3 years- Flash in the pans have the highest and Delayed documents the lowest *CPP/FCSm* scores (even below 1!). When we increase the length of the citation window up to 5 years, Normal documents are the ones with the highest scores, but Flash in the pans still present higher scores in comparison with Delayed documents. If, however, the citation window is increased up to 10 years, Normal documents are still the ones with the highest scores but now Delayed documents outperform Flash in the pans. Finally, with the longest citation window (20 years), Delayed documents show the highest *CPP/FCSm* (mainly above 2!), Normal documents are second in the rank and finally Flash in the pans have the lowest scores. Normal documents exhibit a very similar level of impact in the four figures regardless the citation window used, being always above 1.5 and below 2. This may seem high, but we remind that we discarded all papers with less than 5 external citations, and these papers count for about 50% of all papers with at least 1 external citation. If we would include all papers and also the not cited papers the *CPP/FCSm* value for this whole set would, as expected, be close to 1.

### 3. Evolution of CPP in relation to durability

The impact of documents in relation to their different types of durability has been analyzed. The average 'simple' impact (i.e., impact not normalized by field, such as in case of the *CPP/FCSm*) (*CPP* values) as well as the percentage of the citations received on a yearly basis has been studied, see Fig. 5.

**Figure 5**: Evolution of *CPP* of documents in relation to year of publication

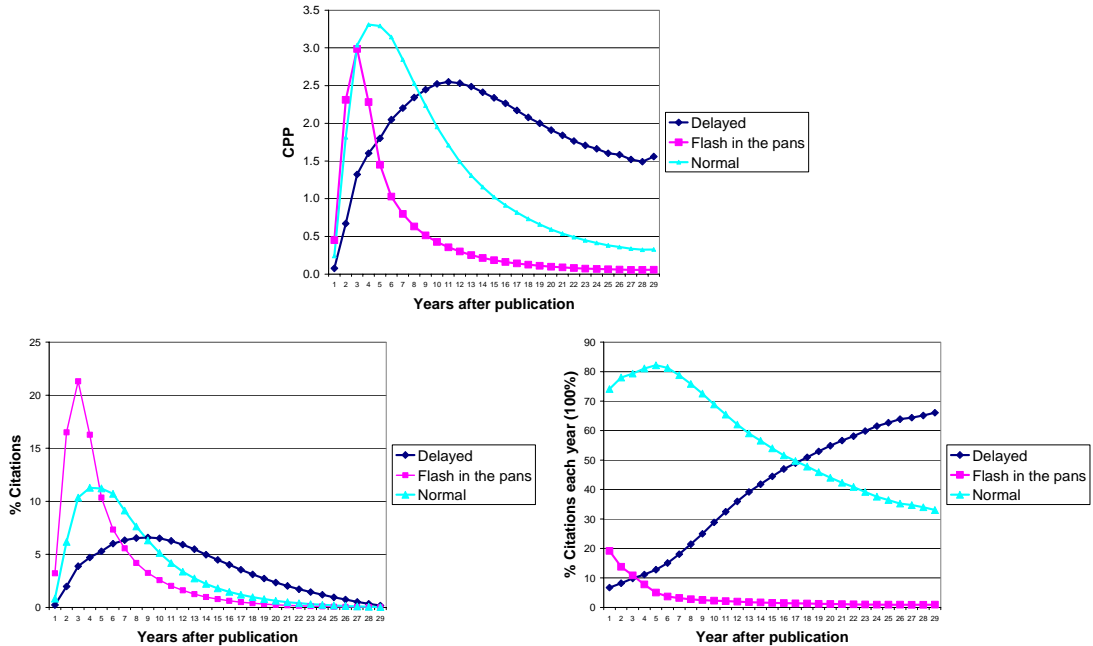

According to Fig. 5 the evolution of the *CPP* presents characteristic trends. All three types of Durability have an increasing pattern at the beginning and after they have reached a peak they start to decline. However, the shapes of these *CPP* patterns are quite different for the three types of durability. Flash in the pans have a very high impact (*CPP*) during the immediate years after publication, reaching their maximum peak after three years and then showing a rapid decline. In contrast, Delayed documents are not very highly cited at the beginning, but their impact increases steadily during several years, reaching a peak after 10-11 years following publication and showing a slow decrease during the rest of the period. It is also interesting to stress that is not until the 5$^{th}$ year that Delayed documents surpass Flash in the pans in *CPP* scores.

Normal documents present a pattern of increasing impact during the first years, reaching a peak in the 4th to 5th year after publication followed by a decrease faster than Delayed documents but slower than Flash in the pans. As the figure in the left bottom-corner shows, Flash in the pans accumulate the highest percentage of their citations during the first years following publication, while Delayed documents have a more even distribution of the percentage of their citations (always below 10%).

Finally, in the right bottom-corner of Fig. 5 we see how the main part of the citations during the first years comes predominantly from Normal documents, but these documents show a decreasing trend after 5-6 years following publication. On the other hand, the percentage of citations represented by Delayed documents increases during the whole period.

Another interesting aspect is a possible country-dependence of the different Durability types. Thus, we analyzed the presence of US institutions in the documents from the perspective of their durability and show the results in Table 9.

**Table 9**: Presence of USA institutions in the documents by Durability

| Final durability | Tot. WoS | Tot USA | %USA | %Non USA |
|---|---|---|---|---|
| Flash in the pan | 814,098 | 364,856 | 44.8 | 55.2 |
| Normal | 5,860,703 | 2,640,768 | 45.1 | 54.9 |
| Delayed | 1,665,712 | 692,852 | 41.6 | 58.4 |
| *Total* | *8,340,513* | *3,698,476* | *44.3* | *55.7* |

We find that Delayed documents have a somewhat lower presence of US institutions because 58.4% of Delayed documents have no US participants (for the total this is 55.7%), while Flash in the pans have a higher US participation as compared to Delayed documents (44.8% vs 41.6%). A possible explanation is that US is generally considered to be more "central" in science and therefore more rapidly noticed and accepted by the scientific community, while ideas from other countries may need more time to be accepted and acknowledged.

**Discussion**

In this paper we present a general methodology for the bibliometric description and classification of documents, describing the phenomenon of durability as measured by their citations. This methodology, and especially the technical definition of durability types, implies an important step in the analysis of the ageing and durability of documents because large sets of documents can be studied (in fact, all documents contained in the *Web of Science* were classified). Moreover, the delimitation of the durability types is more flexible and general than in earlier studies on more extreme cases such as "Sleeping beauties" or "Delayed recognition". Furthermore, the possibility of studying not only highly cited documents but also moderately cited documents implies an important improvement in the analysis of the citation histories of documents, an issue that was suggested before by Aversa (1985).

The methodology discussed in this paper presents several advantages:
- it can be applied to *large sets* of documents;
- it can also be applied to documents published in *different years* (although we consider 5 year of citation history as a minimum threshold for reliable results);
- it takes into account that documents can be classified in *more than one field*;
- it can also be *updated yearly or monthly* and thus improved as more information is available.

Another important advantage of the methodology is its simplicity as only 3 general types of durability are considered, compared to the broader classifications provided by other authors (Avramescu, 1979; Moed et al, 1998). Our approach of three types of durability is also supported by the results of Aksnes (2003) who proposed an initial classification for highly cited papers of 9 different types based on the rise and decline of citations of documents. However, according to the results of this author only three main types account for the larger majority of the papers, being the "Early rise - Rapid decline" (roughly our Flash in the pans), the "Medium rise - Slow decline" (similar to our Normal type) and the "Delayed rise - No decline" (very close to our Delayed type).

The analysis of documents classified by our methodology shows the following main results. By comparing Delayed documents with Flash in the pans, we find that the first receive more citations (**CPP**, which is equal to **C** for an individual document) and that they have a higher field-specific impact (**CPP/FCSm**). This is in agreement with earlier more general results by Aversa (1985), Aksnes (2003) and Levitt and Thelwall (2009) who showed that slower ageing rates and lateness in citations are correlated with higher citation counts.

With respect to the bibliographic properties of the three types of Durability, our first observations necessitate further analysis in follow-up work. To start with, Delayed papers are characterized by, on average, less authors, centers and countries per document than Flash in the pans and Normal documents. These results suggest that Delayed documents are papers with *lower* levels of collaboration. This is an interesting finding because it could imply that less collaborative papers, which are generally considered as papers with lower impact (Glänzel & Schubert, 2001; Persson et al, 2004) appear to be not necessarily of less quality 'in the long run'. In other words, it could be hypothesized that less collaborative papers could also gain high impact but they need more time to become fully recognized and cited accordingly.

A complementary explanation to this idea is that papers in collaboration and especially in international collaboration have larger audiences (van Raan, 1998), thus benefiting from more direct visibility and impact. Also in this line of arguing, papers in collaboration improve their potential visibility by using the network contacts of their collaborators (Katz & Martin, 1997) and thus they can diffuse their findings wider and faster among larger audiences. According to this, less collaborative but still high quality papers (such as Delayed documents) may need more time to be distributed, known and acknowledged by their peers. This idea can also be supported by our finding that Delayed documents also have fewer keywords and references per document. This means that their retrieval through bibliographic databases is more difficult (there are less retrieval elements as compared to Flash in the pans and Normal documents).

On the other side, Flash in the pans are relatively more frequent in the 'shorter' document types (Notes, Letters, Editorials, etc.). This explains why they have the smallest number of pages. This is also linked to the idea that these documents represent an immediacy value rather than an archival value, as they are published in journals that try to present research results as quickly as possible. This is in agreement with the observations of Moed et al (1998) that journals containing rapid communications show a fast decline and 'mature' patterns, particularly journals such as *Nature* and *Science*. In the same line, van Dalen & Henkens (2005) found that Flash in the pans are related with high reputation journals, considering that top journals are the ones where the broader academic debate takes place, while other more specialized journals mostly do not publish work with a broad impact. In any case, these results also point out again the weakness of the Journal Impact Factor as a measure of scientific impact, since it favors journals publishing more Flash in the pans than Delayed documents (even within the same field). This supports the criticism of Seglen (1997) that high impact factor journals are likely covering areas of basic research with a rapid expanding but short-lived literature that use many references per article.

It is also interesting to notice that Flash in the pans are documents published mainly in fields with high citation-density (relatively high **FCSm**), but their overall impact in the long run is much lower than for Delayed documents. Van Dalen & Henkens (2005) also found that Flash in the pans are less cited documents. With the observation that Flash in the pans have more self-citations than Delayed documents, and taking into account that self-citations come mainly during the first years after publication (Glänzel et al, 2004; Schubert et al, 2006), we suggest that Flash in the pans are boosted and advertised by their own authors in the years following publication (Medoff, 2006; Fowler & Aksnes, 2007), thus gaining immediate attention by other researchers, but loosing interest in the long term. In other words, self-citations can boost significantly the impact of documents, because they mainly influence the impact during the first years after publication which are crucially important years for Flash in the pans but not for the more durable papers (i.e. Normal and Delayed documents). Nevertheless, these arguments are based on general self-citation characteristics, as if 'technical' arguments can explain the entire impact history of papers. It is however very well possible that Delayed documents represent research work of another kind than the work published in Flash in the pans and in Normal papers, namely, related to the idea of *Sleeping Beauties*, work that is ahead of time, so that conceptual and intellectual characteristics provide an explanation of the impact history.

Considering the evolution of citations over time for the three types of durability, we observe how Delayed documents present a rise and decline in their impact which is much slower than for the other two types, i.e., the Delayed documents are the more frequently cited documents 10 years after publication. In contrast, Flash in the pans show a very strong growth in impact immediately after publication and then they rapidly decrease after reaching a peak. One possible explanation for this fast declining pattern of Flash in the pans is that they are often 'followed up' by normal papers, which will 'take over' and present the same results but more in detail and more thoroughly treated. Thus, the impact shifts from the Flash in the pan to the normal paper, leaving the role of the Flash in the pans as 'early warnings' in order to get as quick as possible the credits for the results. This explanation reinforces the idea of sets of publications being in fact just 'one oeuvre'. We will focus further research on the identification of such links between of strongly related papers.

With regard to the different citation windows, we find that after 20 years Delayed documents show the highest impact in their fields (surpassing both Normal and Flash in the pans). This finding implies that very short citation windows mainly favor Flash in the pans -and thus it is not surprising the fact that Flash in the pans are published in journals with the highest impact factors- while the opposite is found for Normal and Delayed documents. According to our results it is clear that establishing the percentage of Delayed documents is a crucial topic in order to test if indicators based on shorter citations windows can affect those units of analysis, such as research groups and institutes, with higher degrees of Delayed papers.

Finally, the methodology presented in this paper raises many questions about the ageing of documents and how documents can be considered for bibliometric analysis. Several intriguing lines of future research are suggested, among them we can mention:

- Better understanding of the three general types of durability, *improving their characterization* and providing more explanations for the general patterns found in this paper, as well as the consideration of other aspects of interdisciplinarity, self-citations, etc.

- Study of *different scientific fields* in order to detect main differences among them; also deviations from the general pattern according to durability is an important issue of future research.

- Analysis of the *effects* of these durability types for the calculation of general *bibliometric indicators* is very important as well as the establishment of more appropriate citation windows.

- Importance of the durability types for the study of *different units of analysis* and aggregation levels is highly important in order to find out whether the impact assessment of groups, institutes or even of individual researchers can be affected by these types of documents. In this line, important questions are how these durability types are distributed among different institutes and who are the researchers behind these types; these issues are directly related to a better understanding of the ageing of scientific literature.

- Identification of strong links between papers with different classification, particular, as discusses above, between Flash in the pans and Normal papers;

- Finally, the *prediction of impact* is an important topic of research nowadays (Geller et al, 1986; Feitelson & Yovel, 2004; Mingers & Burrel, 2006; Castillo et al, 2006).

These issues illustrate the potential for further research of the developed methodology.